# Impact arising from sustained public engagement: A measured increase in learning outcomes

*James A McLaughlin [1], Lynda G Boothroyd [2] & Peter M Philipson [1]*

1 = Northumbria University, Newcastle upon Tyne, UK
2 = Durham University, County Durham, UK

## Abstract

This article details the impact arising from a sustained public engagement activity with sixth-form students (16 to 17 year olds) across two Further Education Colleges during 2012/13. Measuring the impact of public engagement is notoriously difficult. As such, the engagement programme closely followed the recommendations of the National Co-ordinating Centre for Public Engagement (NCCPE) and their guidance for assessing Research Excellence Framework 2014 (REF2014) impact arising from public engagement with research. The programme resulted in multiple impacts as defined by the REF2014 under "*Impacts on society, culture and creativity*". Specifically:

- the beneficiaries' interest in science was stimulated;
- the beneficiaries' engagement in science was improved;
- their science-related education was enhanced;
- the outreach programme made the participants excited about the science topics covered;
- the beneficiaries' awareness and understanding was improved by engaging them with the research;
- tentative evidence of an improvement in AS-level grades;
- indirect evidence of an improvement in student retention.

These impacts were evidenced by the user feedback (i.e. sixth-form students) collected from 50 questionnaires (split 16 and 34 across the two Further Education Colleges), as well as testimonies from both the teachers and individual participants.

This article will be of interest to anyone looking at how to evidence that public engagement has produced impact, in particular with regards to impact arising from a sustained public-engagement activity.

**Key messages**

- Ongoing interactions with the same group of people multiple times over a fixed period leads to impacts in their motivation, understanding and awareness.
- This case study outlines an approach to designing and delivering sustained public-engagement activity.
- This case study describes how evidence of impact was gathered and evaluated.

Corresponding author
James A McLaughlin, Department of Mathematics, Physics and Electrical Engineering,
Northumbria University, Newcastle upon Tyne, NE1 8ST, UK.
Email: james.a.mclaughlin@northumbria.ac.uk

# Introduction

The 2014 Research Excellence Framework, REF2014, was a system for assessing the quality of research in higher education institutions (HEIs) in the UK, replacing the Research Assessment Exercise, RAE, which was conducted in 2008. HEIs were invited to make submissions by 29 November 2013 to one of thirty-six units of assessment. Comprehensive guidance regarding the assessment criteria was detailed in the REF2014 Panel Criteria and Working Methods document (REF, 2014).

REF2014 contained a new submission criterion not seen in the RAE: *Impact*. Specifically, impact entered the REF2014 assessment via the REF3a document, a template describing the submitted unit's approach during the impact assessment period (1 January 2008 to 31 July 2013) to enabling impact from its research, and via the REF3b document, which detailed case studies as specific examples of impacts achieved during the assessment period, underpinned by research conducted by that institution within the period 1 January 1993 to 31 December 2013.

Impact, which can be understood as the *non-academic benefit* of research, can take a range of forms, and the REF2014 panels welcomed case studies describing impacts that have provided benefits to, for example culture, the economy, the environment, health, public policy and services, quality of life, and society. Impacts could also be felt locally, regionally, nationally or internationally. Impact case studies were assessed on the 'reach' and 'significance' of the impact in the above areas.

One such example of impacts were those arising from public engagement activity, for example as described on page 51 of the REF2014 Panel Criteria and Working Methods (REF, 2014). The REF2014 framework document (paragraph 161) explicitly encouraged impacts arising from engaging the public with research to be featured in REF2014 submissions:

> *"There are many ways in which research may have underpinned impact, including but not limited to: (c) Impacts on, for example, public awareness, attitudes, understanding or behaviour that arose from engaging the public with research. In these cases, the submitting unit must show that the engagement activity was, at least in part, based on the submitted unit's research and drew materially and distinctly upon it".*

This article details a <u>sustained</u> public engagement activity, i.e. ongoing interactions with the same group multiple times over a fixed period, as well as the evidenced <u>impact</u> arising from the activity, as defined in the REF2014 guidance as "*Impacts on society, culture and creativity*".

Public engagement with science has gained widespread support from UK and EU science in recent years (e.g. House of Lords Select Committee on Science and Technology, 2000; Department for Innovation, Universities and Skills, 2008; Research Councils UK, 2010) and such engagement can occur in a variety of formal or informal settings (e.g. MacNaghten et al., 2005; Stilgoe et al., 2006; Holliman & Jensen, 2009; Jensen & Holliman, 2009; and references therein), including public lectures and science festivals (e.g. Holliman et al. 2009; Jensen & Buckley, 2012, and references therein) and universities (e.g. Grand et al., 2015). Impact arising from public engagement is well documented in journals such as *Public Understanding of Science* (e.g. see article by, and special issue introduced by, Stilgoe et al., 2014; and references therein) and, moving forward, *Research For All* (e.g. Duncan & Oliver, 2017). Thus, let us be clear about the purpose of this current article: this article details impact arising from *sustained* public engagement, i.e. ongoing interactions with the *same* group multiple times over a fixed period, and evidences that impact.

**Underpinning research and outreach programme**

A key feature to all types of REF2014 impact was that for the November 2013 submission, the impact must have been experienced within a particular period (1 January 2008 to 31 July 2013) and that it must have been *underpinned by research* conducted at the submitting institution within the period 1 January 1993 to 31 December 2013.

Professor McLaughlin (hereafter referred to as McLaughlin) was the principal investigator of this sustained public engagement activity and joined Northumbria University in January 2010. His research interests include investigating magnetic fluid dynamics and mathematical modelling of solar and astrophysical processes. The public engagement activity was underpinned by specific research carried out by McLaughlin at Northumbria University, i.e. carried out at the submitting institution. With this in mind, we now describe the research underpinning the public engagement activity (published by McLaughlin in the period January 2010 – December 2013) followed by the construction of the programme itself.

In 2011, McLaughlin had published a detailed review article on magnetohydrodynamic (MHD) wave behaviour within inhomogeneous magnetic media (McLaughlin et al., 2011). Morton, Verth, Erdélyi & McLaughlin (2012) reported a novel application of *magnetoseismology* – the application of MHD wave theory to magnetic wave observations to probe the plasma – to describe the properties of a previously unseen dark thread accompanying a solar jet. McLaughlin et al. (2012a) reported new results from oscillatory reconnection (time-dependent, wave-generating reconnection) which demonstrated that oscillatory reconnection driven by magnetic flux emergence provides a natural explanation for generating the observed (transverse) solar jets, and McLaughlin et al. (2012b) investigated the sensitivity of the reconnection mechanism to various parameters. Finally, Morton & McLaughlin (2013) analysed high-resolution observations from state-of-the-art solar satellites.

In 2012, McLaughlin constructed a five-part outreach programme of presentations: (i) *Introduction to the Sun*; (ii) *The Sun and its effect on the Earth and Space Travel*; (iii) *Electromagnetism and MHD*; (iv) *Special Relativity*; and (v) *Mars*. Key aspects of the presentations were based on the specific research carried out within the unit by McLaughlin. For example:
- Outreach materials covering magnetic flux emergence and magnetic reconnection were underpinned by research detailed in McLaughlin et al. (2012a) and (2012b), respectively.
- Outreach materials explaining solar observations were underpinned by research reported in Morton & McLaughlin (2013).
- A detailed mathematical model of the solar wind was presented as part of Presentations (ii) and (v), and this model was developed as part of US Air Force grant FA8655-13-1-3067 (McLaughlin, 2013).
- Presentation (iii) contained a discussion of MHD waves underpinned by research from McLaughlin et al. (2011; 2012b).

The rest of the outreach materials were created from McLaughlin's body of research carried out within the submitting institution. The outreach programme was also supported by Higher Education Innovation Funds (McLaughlin, 2012) to cover travel costs. The cited outputs above are specific examples of key research outputs produced within the institution. These outputs detail the original research, whereas the outreach programme acted as the vehicle to engage the audience with the research.

Thus, it is clear that the public engagement programme was underpinned by research carried out *within the submitting unit of assessment* and *within the permitted period*.

## Overall approach to evaluation

### Early planning

The activity was specifically designed to follow closely the recommendations of the National Co-ordinating Centre for Public Engagement, NCCPE, and their guidance for assessing REF2014 impact arising from public engagement with research (NCCPE, 2011a). The entire outreach activity was built specifically with evidence gathering in mind, in order to be certain to satisfy the NCCPE guidance once the activity was completed. What really drew our attention was the NCCPE's comment with regards to *sustained* public engagement activity, i.e. ongoing interactions with the same group multiple times over a fixed period (see "Pilot study" below).

In addition to following the guidance from NCCPE (2011a), our approach was influenced by three main factors:
- A 'pilot study' – see below – which helped shape the questionnaire and explored evidence gathering approaches. These early stages helped plan the extended, sustained outreach programme.
- Previous public outreach lectures – delivered by McLaughlin at schools, National Science Weeks, Space Week (in USA) – informed the optimum duration of information delivery to 16 and 17 year old school students.
- Presentations on best practice for science-related public outreach at National Astronomy Meetings and via follow-up meetings with the Royal Astronomical Society. McLaughlin has been heavily involved in discussions with the Royal Astronomical Society on evidence-gathering mechanisms for Astronomy-related public engagement.

### Pilot study

In March 2012, McLaughlin presented his research to a general audience as part of the Newcastle Science Festival. At that festival, McLaughlin interviewed a subset of the audience after his presentation, using a questionnaire. A subset was chosen due to time-constraints (McLaughlin asked the audience if people wouldn't mind staying on after his presentation, and only a subset agreed). In the interviews, McLaughlin was hoping to learn if the audience found the presentation interesting, engaging, exciting, and how the clarity of the presentation could be improved for future sessions. He realised that such interviews, or more specifically that evaluation data and user feedback, could also be used to measure changes in knowledge and behaviour. However, it was also recognised that a one-off presentation might influence audience members on the immediate timescale, but a *sustained* programme with the same group was needed to embed a lasting benefit. The questionnaire used during the Newcastle Science Festival was adapted and improved into the version used for the extended, sustained public outreach programme. In this way, the pilot study clearly informed what we did for the sustained outreach programme.

## Planning and Organising the Outreach programme

### Initial Contact

The activity was planned by the researchers and Further Education college teachers. To begin organising the outreach programme, several Further Education colleges were contacted to see if they were interested in participating. We discussed the length of the outreach programme (October to June) and the overall concept. The researchers would create and deliver the outreach material (with this college teachers guidance – see below), and in return asked the Further Education college teachers to agree to allocate time for five individual sessions throughout the academic year. This was a challenging request, since the academic year is naturally busy at Further Education colleges, but the potential benefits (improved interest and engagement in science, as well as strengthening links to universities) were also made clear. Five sessions were chosen as a compromise between a desire to have multiple interactive sessions (to achieve the *sustained* nature of the activity) versus available time in the Further Education colleges academic calendar.

**Creating the Materials and Teacher Input**

The researchers created the outreach material, and planned the delivery of material of increasing complexity over five distinct presentations. The teachers gave guidance on the AS-level Physics syllabus and, together, the researchers and teachers planned the timetabled delivery of the material over the duration of a single school academic year. The presentations were designed with the AS-level Physics syllabus in mind, for example, the syllabus material on magnetism – including specifically *when* in the year that section is delivered – were directly related to presentation (iii) *Electromagnetism and MHD*. In this way, each presentation acted in synergy with the AS-level material. In other words, the activity was planned to support teachers to use astronomy/solar physics-related material as a context for curriculum teaching. This synergy was also important in helping secure agreement from the Further Education college teachers to host multiple sessions, since this aligned with the college teachers' primary objective to deliver the syllabus material.

The researchers and teachers also discussed the presentations immediately after (in person) and between sessions (via email) in order to improve subsequent ones. In practice, this did not result in any real changes to the programme (since this had already been discussed in the initial contact) but the option was there.

**Choice of Session Duration**

Feedback from previous public engagement activities with school students aged 16 and 17 also informed how the programme of activities was planned. School students gave feedback to the teachers prior to the activities. McLaughlin is passionate about public outreach, and for many years has given public lectures about his research. He presents to all age groups. He has also visited schools and presented to 16 and 17 years olds and, in his experience, has found that 20 minute oral presentation are more effective than, say, university-style 60 minute lectures. This past experience shaped the sustained outreach programme, such that each of the five presentation sessions was designed to last for one hour, and was divided into a 20 minute oral presentation followed by 5-10 minutes of questions, followed by a second 20 minute session, again followed by 5-10 minutes of questions. This format was judged most appropriate for the target audience of 16 to 17 year olds, and this planning was informed by feedback from the teachers and the school students.

**Structure and Timing of the Activity**

We devised a multiple visit, ongoing public outreach programme where McLaughlin interacted with the *same* group five times in a structured set of exercises. The target audience was the first year of AS-level Physics, since this has a naturally alignment to McLaughlin's research. The aim was to investigate the positive effects and benefits of public engagement at these sixth-form colleges using outreach materials underpinned by McLaughlin's research into Solar Physics and Magnetohydrodynamics (MHD). We worked with two Further Education colleges in the collaborative outreach programme (the two locations are anonymised in this paper, and referred to as College One and College Two). Throughout their studies, the students had attended occasional one-off public engagement activities, but (according to the teachers and students) they had never been involved in an ongoing, multiple-visit programme.

Between October 2012 and June 2013 (2012/13 school year), McLaughlin visited the AS-level Physics students at these two sixth-form colleges on a total of ten independent occasions and at each college delivered five individual presentations (of increasing complexity). The five individual presentations were on the following topics: (i) *Introduction to the Sun*; (ii) *The Sun and its effect on the Earth and Space Travel*; (iii) *Electromagnetism and MHD*; (iv) *Special Relativity*; and (v) *Mars*.

It was believed that the information, knowledge and benefits of the outreach programme take time to really 'sink in' and so an outreach programme spread across a whole school year seemed an appropriate length of time. This is related to the idea of deep learning in education theory (see, e.g., Biggs, 1999; Entwistle, 1988; Ramsden, 1992; Case and Marshall, 2004; Houghton, 2004; and references therein). Moreover, McLaughlin's research (which underpinned the outreach material) is specialist material, and it took time to build the knowledge and context for the audience to a mature level in order to engage properly with and understand the underpinning research. Simply put, it would not be appropriate to launch straight into the details; which may have inhibited the audience's understanding.

**Evidence gathering**

Evaluation questionnaires were completed by the participants at the end of the fifth presentation and an analysis of the responses was performed. The questionnaire (see Appendix) was constructed using the Likert scale method, with options for free text and open questions. There were seven Likert scale questions covering their self-evaluated interest and engagement with science in relation to the programme. The Likert scale questions were worded specifically to assess the examples of impact on society, culture and creativity from the REF2014 guidance (see "Assessment criteria: impact" under REF, 2014).

The students signed up for the programme themselves (sign up was not compulsory) though, naturally, the sessions were promoted by the teachers. Registers were taken at each of the five presentations to ensure that students completing the questionnaires had indeed attended the whole outreach programme. This resulted in 50 completed questionnaires. Hence, these 50 beneficiaries represent a subset of the total number of students studying AS-level Physics across both colleges. For example, a student who attended, say, only four presentations was excluded from completing the questionnaire. Non-attendance was assumed to be unrelated to the study, i.e. students who missed sessions did so independently of the programme. Specifically, College One had 58 students enrolled on AS-level Physics and, at the end of the activity, questionnaires were completed by 34 students (59%) and College Two had 36 students and questionnaires were completed by 16 students (44%). A hardcopy of the questionnaire was given out to the students (at the end of the fifth session), was completed by the students and then was collected back within the same session.

Data from the 50 evaluation questionnaires was used to evidence the impact on the change in participants' interest, knowledge, engagement and motivation. There was 100% response rate from all 50 participants who individually each answered 100% of the questions. 92% of participants rated the overall outreach programme as good or very good (38%=very good; 54%=good; 8%=average; 0%=poor; 0%=very poor).

## Impact

The public engagement programme resulted in multiple impacts as defined by the REF2014 guidance as "*Impacts on society, culture and creativity*". Specifically:
- the beneficiaries' interest in science was stimulated,
- the beneficiaries' engagement with science was improved,
- their science-related education was enhanced,
- the outreach programme made the participants excited about the science topics covered,
- the awareness and understanding of the beneficiaries was improved by engaging them with the research,
- tentative evidence of an improvement in AS-level grades,
- indirect evidence of an improvement in student retention.

These impacts were evidenced by questionnaire responses (user feedback), participants' quotes, AS-results and testimonies. A summary breakdown of the feedback survey is given below.

## Specific impacts and specific evidence

The impact "*the beneficiaries' interest in science was stimulated*" was evidenced by the results of two feedback questions:
- **As a direct result of Dr McLaughlin's outreach programme, are you now more interested in science as a subject than you were before?**
- ➢ 54% of responses indicated that they were more (or much more) interested in science as a subject than they were before, as a direct result of Dr McLaughlin's outreach programme (10%=much more; 44%=more; 44%=about the same; 2% less; 0% much less).
- **Please indicate to what extent you agree or disagree with the following statement: "My interest in science has been stimulated as a direct result of Dr McLaughlin's outreach programme".**
- ➢ 64% of responses agreed (or strongly agreed) with this statement (14%=strongly agreed; 50%=agreed; 26%=neither agree nor disagree; 8% disagree; 2% strongly disagree).

The impact "*the beneficiaries' engagement with science was stimulated*" was evidenced by the results of two feedback questions:
- **As a direct result of Dr McLaughlin's outreach programme, are you now more likely to talk to your teacher about science?**
- ➢ 32% of responses indicated that they were more (or much more) likely to talk to their teacher about science than they were before (2%=much more; 30%=more; 68%=about the same; 0%=less; 0%=much less).
- **As a direct result of Dr McLaughlin's outreach programme, are you now more likely to consider studying science at university?**
- ➢ 34% of responses indicated that they were more (or much more) likely to consider studying science at university than they were before (6%=not planning on going to university; 12%=much more; 22%=more; 58%=about the same; 2% less; 0% much less).

The impact "*their science-related education was enhanced*" was evidenced by the results of the following feedback question:
- **Please indicate to what extent you agree or disagree with the following statement: "My science-based education has been enhanced as a direct result of Dr McLaughlin's outreach programme".**

- 68% of responses agreed (or strongly agreed) with this statement (4%=strongly agreed; 64%=agreed; 26%=neither agree nor disagree; 6% disagree; 0% strongly disagree).

The impact "*the outreach programme made the participants excited about the science topics covered*" was evidenced by the results of the following feedback question:
- **Did the outreach programme make you excited about the science topics covered?**
- 80% of responses indicated yes (or yes to a strong extent) in answer to this question (10%=Yes to a strong extent; 70%=yes; 18%=not sure; 0%=no; 2%=no to a strong extent).

The impact "*The awareness and understanding of the beneficiaries was informed, by engaging them with the research*" was evidenced by the following illustrative feedback from the participants, i.e. participants' quotes from survey (quotations may also evidence the other impacts):
- "*It has extended my knowledge of science and gave me more motivation to do well in science. I am much more enthusiastic about it now*".
- "*It did make me more enthusiastic about physics, discovering new things, and realising that there are so many things that you don't know*".
- "*I realise that I can research topics myself in order to increase my knowledge*".
- "*It's given me more knowledge on the subjects discussed. It's made me more interested in learning physics in more detail, rather than just what is learnt in lessons*".
- "*Increase general interest in topics I didn't even know existed*".
- "*I feel more happy going into A2 with a higher knowledge about space. A lot of the topics discussed I wasn't confident* [on] *beforehand*".
- "*Has given me more enthusiasm for science and I would like to learn more*".
- "*I would like to take geology at university and am now going to look at the courses to see if they include geology on other planets*".
- "*It inspired me to do a space based EPQ (about Mars)*". [Extended Project Qualifications are part of level three of the National Qualifications Framework].

**Additional Significance**

There is additional evidence that the outreach programme also resulted in an improvement of AS-level grades. A testimony from the Head of Physics at College One states:
> "*The topics covered encouraged several students with a lower than average ALIS* [Advanced Level Information System, ALIS 2013] *predicted grade to attend the presentations. Engaging some of these students was an achievement in itself and from the discussions that followed it seemed to have a motivational effect. Statistically speaking these students have a lower than average chance of achieving the high grades required in order to gain a place at university making engaging and motivating them to learn even more important*".

In addition, a comparison of the AS-level Physics predicted grades versus actual grades across the College One students showed a clear increase in grades, specifically:
- **Predicted grades**:   A= 0%, B=18%, C=23%, D=14%, E=45%
- **Actual grades**:      A=18%, B=27%, C=18%, D=9%, E=27%

with an average increase of +0.86 grade per students (not uniform increase).

Thus, from the testimony and grade comparison, there is tentative evidence that the programme contributed to an improvement in AS-level grades, i.e. improved attainment.

Furthermore, there is some (indirect) evidence of increased retention and progression. A testimony from the Head of Physics at College Two states:
> "*This year 60% of students have continued from AS level Physics to A2 level Physics compared to an average of 50% over the last few years. I would not be able to hold Dr*

> *McLaughlin completely responsible for this, but do believe that his delivery of lectures through the year has indeed partially contributed to this success".*

**Additional Reach**

The target audience was the first year of sixth form of AS-level Physics. Across the two colleges, data was collected from 50 AS-level students (16-17 year olds), consisting of 39 males and 11 females where, for STEM subjects, NCCPE (2011a) defines female students as a 'hard to reach' audience.

**Institutionalising the programme within the University**

Within Northumbria University, our activities have helped shape our STEM-related public engagement activities. Specifically, the University has a dedicated physics and astrophysics outreach centre called *Think Physics* (Think Physics, 2014) which highlights the benefits of STEM to young people and their key influencers, including parents and teachers. This £1.2 million HEFCE Catalyst-funded project (November 2014 - December 2017) aimed to increase the uptake of physics and related disciplines, with a focus on growing science capital among women and other under-represented groups. Think Physics works across all age-ranges and works in partnership with industry, Local Education Authorities, schools, museums, education and science trusts.

Part of the *Think Physics* activity and interaction with schools involve an aspect of evaluation and evidence gathering. Our activities have directly influenced this HEFCE Catalyst-funded project such that data is now being collected both pre- and post-activity, combined with follow-up data collection downstream to assess longer-term effects, changes and/or benefits, e.g. Padwick et al. (2016). McLaughlin was involved in writing the successful *Think Physics* bid to HEFCE, hence was in a position to reflect upon his public engagement activities with school students whilst writing the bid. The *Think Physics* programme ran from 2014 to 2017, and has now evolved into NUSTEM (NUSTEM, 2018).

**Conclusion**

We have detailed the impact arising from a sustained public engagement activity with sixth-form students across two Further Education Colleges over the period October 2012 to June 2013. The target audience was the first year of sixth form of AS-level Physics. Across the two colleges, data was collected from 50 AS-level students (16-17 year olds), consisting of 39 males and 11 females. The programme resulted in multiple impacts as defined by REF2014 under "*Impacts on society, culture and creativity*". Specifically: the beneficiaries' interest in science was stimulated; the beneficiaries' engagement in science was improved; their science-related education was enhanced; the outreach programme made the participants excited about the science topics covered; the beneficiaries' awareness and understanding was improved by engaging them with the research; and (tentative/indirect) evidence of an improvement in AS-level grades and in student retention. These impacts were evidenced by the user feedback collected from 50 questionnaires, and testimonies from the teachers and individual participants.

We believe this article will be of interest to:
- Anyone interested in how to demonstrate and evidence that public engagement has produced impact (e.g. changes in behaviour, process, policy, etc).
- Anyone interested in impact arising from a *sustained* public engagement activity, i.e. ongoing interactions with the same group multiple times over a fixed period.

Impact was a new submission criterion in REF2014 and will form part of REF2021 (REF, 2021). Impact also forms a key part of many Research Council UK (RCUK) grant applications, for example through the '*Pathways to Impact*' section. Impact requires a demonstration that there has indeed been a change (in behaviour, process, policy, etc) and so a crucial part of this is gathering suitable evidence. This article details our public engagement activity as well as our specific evidence-gathering approach and we believe this is a useful blueprint and example of best practice in evidencing impact arising from *sustained* public engagement. We have explained clearly our methodology, as well as our approach to increasing the 'reach' and 'significance' of the impact. We have also following closely the recommendations of the National Co-ordinating Centre for Public Engagement and their guidance for assessing REF2014 impact arising from public engagement with research.

We also envisage that our article will be of great interest to the scientific community who are specifically interested in linking their research to public engagement, for example as part of their RCUK '*Pathways to Impact*' documentation.

For individuals or institutions interested in our approach, we also suggest potential improvements to the programme. Firstly, we did not analyse whether there was any impact on those who only participated in some of the sessions (due to our approach of ensuring that students completing the questionnaires had attended all five sessions of the outreach programme) but, had we done so (e.g. distribute questionnaires to all students on the AS-level and get them to state how many session they had attended) then potentially this could be a useful way of assessing whether the benefits arise to those attending the full set of sessions or perhaps even 'how many' sessions are needed for an (optimal) sustained engagement. Secondly, the Higher Education Access Tracker service (HEAT, 2018) was not available at the time of this study, but could be used in future programmes to track what students go on to do after their AS-levels.

Finally, we note that with the introduction of the impact submission criterion to REF2014 and its inclusion in REF2021, there is a real risk that all public engagement may become focused on the REF impact method of assessment, meaning that other valuable forms of engagement will not be supported or valued. We hope that this risk, as well as the definition of the impact of engagement, is challenged and debated within the pages of *Research For All*.

## Acknowledgements

JAM acknowledges STFC for support via grant numbers ST/L006243/1 and ST/N005562/1.

**Appendix: Feedback Questionnaire**

**FEEDBACK QUESTIONNAIRE : Outreach Programme**

Feedback is a vital part of our programme development and quality improvement. This is your chance to air your views anonymously. Please take time to give us your honest and constructive comments on your experiences.

**How interested were you in science as a subject before Dr McLaughlin's outreach programme began?**

☐ Very interested

☐ Interested

☐ Don't remember

☐ Not interested

☐ Absolutely no interest

**As a direct result of Dr McLaughlin's outreach programme, are you now more interested in science as a subject than you were before?**

☐ Much more

☐ More

☐ About the same

☐ Less

☐ Much less

**Did the outreach programme make you excited about the science topics covered?**

☐ Yes, to a strong extent

☐ Yes

☐ Not sure

☐ No

☐ No, to a strong extent

**As a direct result of Dr McLaughlin's outreach programme, are you now more likely to talk to your teacher about science?**

☐ Much more likely

☐ More likely

☐ About the same / No change

☐ Less likely

☐ Much less likely

**As a direct result of Dr McLaughlin's outreach programme, are you now more likely to consider studying science at university?**

☐ Not planning on going to university

☐ Much more likely

☐ More likely

☐ About the same

☐ Less likely

☐ Much less likely

**Please indicate to what extent you agree or disagree with the following statement:**

**"My interest in science has been stimulated as a direct result of Dr McLaughlin's outreach programme".**

☐ Strongly agree

☐ Agree

☐ Neither agree nor disagree

☐ Disagree

☐ Strongly disagree

**Please indicate to what extent you agree or disagree with the following statement:**

**"My science-based education has been enhanced as a direct result of Dr McLaughlin's outreach programme".**

☐ Strongly agree

☐ Agree

☐ Neither agree nor disagree

☐ Disagree

☐ Strongly disagree

**Please give your overall view of Dr McLaughlin's outreach programme?**

☐ Very good

☐ Good

☐ Average

☐ Poor

☐ Very poor

**What have you enjoyed most about the outreach programme?**

**Has the outreach programme influenced you in any other ways that you would like to mention? e.g. your behaviour, knowledge, enthusiasm, motivation, creativity or in any other activities.
If so, please give details:**

If you are happy to be contacted to discuss your answers further, please write your email address below.

**Do you have any further comments you would like to make on any aspect of Dr McLaughlin's outreach programme?**